\documentclass[apjl]{emulateapj}
\usepackage{epstopdf}
\usepackage{natbib}
\usepackage{amsmath}
\usepackage{enumerate}
\begin{document}
\title{A Simple Physical Model for the Gas Distribution in Galaxy Clusters}

\author{Anna Patej\altaffilmark{1} and Abraham Loeb\altaffilmark{2}}
\altaffiltext{1}{Department of Physics, Harvard University, 17 Oxford St., Cambridge, MA 02138}
\altaffiltext{2}{Department of Astronomy, Harvard University, 60 Garden St., Cambridge, MA 02138}
\email{Electronic Address: apatej@physics.harvard.edu}

\begin{abstract}

The dominant baryonic component of galaxy clusters is hot gas whose distribution is commonly probed through X-ray emission arising from thermal bremsstrahlung. The density profile thus obtained has been traditionally modeled with a $\beta$-profile, a simple function with only three parameters. However, this model is known to be insufficient for characterizing the range of cluster gas distributions, and attempts to rectify this shortcoming typically introduce additional parameters to increase the fitting flexibility.  We use cosmological and physical considerations to obtain a family of profiles for the gas with fewer parameters than the $\beta$-model but which better accounts for observed gas profiles over wide radial intervals.

\end{abstract}


\section{Introduction}\label{s:intro}
Gas hotter than $10^7$ K permeates the intergalactic media of galaxy clusters and constitutes the dominant baryonic component of these virialized objects. Understanding the properties and distribution of this intracluster gas yields insights into cosmology and the physics of galaxy clusters. For example, knowledge of the density and temperature profiles of the gas from X-ray data, coupled with the assumption of hydrostatic equilibrium, provides an estimate of the total mass of the cluster, which can be used statistically to constrain cosmological parameters \citep{av2009b,ettori_review}. High-resolution hydrodynamical simulations incorporate the physical phenomena -- supernova, AGN feedback, as well as radiative cooling -- that shape the observed density and temperature profiles of galaxy clusters \citep[see][and references therein]{kravtsov_review}.

At the intersection of the observational and theoretical fronts lies analytical modeling of the density profiles of X-ray emitting intracluster gas. The gas density has been described by a variety of functions, most notably the $\beta$-model \citep{cavaliere76} and modifications thereof \citep{av2006}. Such functions are either ad hoc models with many parameters to provide fitting flexibility or motivated using unfounded physical assumptions (the popular $\beta$-model, for example, is derived assuming the isothermality of the cluster gas). More recent works have attempted to establish physically motivated models via the assumptions of hydrostatic equilibrium and, for instance, a polytropic equation of state \citep{ostriker05,bulbul2010} or an analytical temperature profile tailored to data \citep{ascasibar08}. However, these models still typically require more parameters than the simple $\beta$-profile to model the range of cluster features observed in X-ray data and impose quite stringent constraints on the physics of the gas. Our goal is to provide a prescription for a new family of profiles that satisfies a set of simple physical assumptions and that has sufficient fitting flexibility to improve upon -- but as many or fewer parameters as -- the standard $\beta$-model.

\section{Background}
\subsection{Gas Density Profiles}
Galaxy cluster gas density profiles have been traditionally described by the three-parameter $\beta$-model,
\begin{align}
\rho_{\mathrm{g}}(r)=\frac{\rho_0}{\left(1+(r/r_c)^2\right)^{3\beta/2}},
\end{align}
with a standard value of $\beta=2/3$. The $\beta$-model is physically motivated; that is, it may be derived from the equation of hydrostatic equilibrium assuming that the gas is isothermal and that the total matter distribution is described by a King model \citep{cavaliere76}. However, the cluster gas is not isothermal \citep{av2005,pratt07,leccardi08}. Furthermore, the $\beta$-model is limited in its ability to describe the observed features of galaxy cluster gas density distributions. Many clusters, particularly the relaxed clusters most often targeted for X-ray measurements, feature cuspy profiles, while the $\beta$-model produces a central core, which is more typical of merging systems~\citep{av2006}. Accordingly, analyses relying on the $\beta$-model often exclude data in the centers out to large radii in order to obtain good fits \citep[e.g.,][]{point04}. 

To circumvent this issue, many papers have built upon the $\beta$-model by adding more parameters to increase modeling flexibility. One such model is the double $\beta$-model \citep[e.g.,][]{mohr1999}, which simply adds an additional $\beta$-profile with a different core radius and $\beta$ parameter to separately fit the inner and outer regions. Another, more recent model is that of \citet{av2006}, who modified a $\beta$-model base by adding functional features to match observations, yielding a ten-parameter model:

\begin{multline}
\rho_{\mathrm{g}}^2(r)=\frac{\rho_{0,1}^2}{(1+(r/r_s)^{\gamma})^{\epsilon/\gamma}}\frac{(r/r_{c,1})^{-\alpha}}{\left(1+(r/r_{c,1})^2\right)^{3\beta_1-\alpha/2}}\\+\frac{\rho_{0,2}^2}{\left(1+(r/r_{c,2})^2\right)^{3\beta_2}}.
\end{multline}
However, although such models greatly improve the fit to data, their forms are no longer motivated by physical principles but are rather constructed from knowledge of variations in cluster gas profiles. Our approach is different; rather than starting with a knowledge of gas profiles, we begin with a few physical and cosmological requirements, in particular that the virialized baryon content of the cluster is representative of the universe and that the cluster features a virial shock.

\subsection{The Virial Shock}
Cosmological theories of structure formation predict that matter accreting onto galaxy clusters should experience a shock, termed the `virial shock,' coinciding roughly with the virial radius of the cluster \citep[e.g.,][]{bertschinger85}. Detection of such a shock around a cluster would serve as an important probe of cosmological infall and accretion. Previous work has examined the impact of shocks on relations between X-ray observables such as the luminosity-temperature relation \citep{cavaliere97}, but we will instead relate the physics of the shock and the dark matter content of the cluster in deriving families of gas density profiles. 

To do so, we will rely on the Rankine-Hugoniot shock jump conditions, which, for the gas density, state that \citep{landau}:
\begin{align}
\Gamma_\mathrm{g}\equiv\frac{\rho_2}{\rho_1}=\frac{(\gamma+1){\cal{M}}^2}{(\gamma-1){\cal{M}}^2+2},
\end{align}
where $\gamma$ is the gas adiabatic index, $\cal{M}$ denotes the shock Mach number, and $\rho_2$  and $\rho_1$ are the densities of the cluster gas and the infalling gas, respectively, at the shock. For a monatomic gas with $\gamma=5/3$, the maximum value of the jump is $\Gamma_{\mathrm{g}}=4$ as ${\cal{M}}\rightarrow\infty$. 

\citet{diemer14} have recently found that dark matter profiles in fact exhibit a similar jump near the virial radius of the cluster \citep[see also][]{adhikari14} . In analogy to the gas shock jump parameter $\Gamma_\mathrm{g}$, we will parameterize the jump in the dark matter density by $\Gamma_\mathrm{DM}$.

\section{Method}\label{s:method}
We seek to obtain gas density profiles not from an ad hoc prescription tailored to suit a data set, but rather from making only a few simple assumptions about the structure and physics of the galaxy cluster, including the virial density jumps discussed above. As shall be explored further in $\S$\ref{s:obscomp}, the result is a family of profiles with up to as many parameters as the $\beta$-model, but better able to accommodate the diversity of features observed in cluster gas profiles.

\subsection{Assumptions}\label{s:assumptions}
Our model galaxy cluster is spherically symmetric, with a virial shock occurring at a radius $s$ and a known dark matter profile $\rho_{\mathrm{DM}}(r)$. We define $f_g=\eta\Omega_b/\Omega_{DM}$, where $\eta$ is a parameter fixing the fraction of baryons in the gas phase within $s$, and $\Omega_b$ and $\Omega_{DM}$ are the cosmological density parameters of baryons and dark matter, respectively. We then assume the following three conditions:

\begin{enumerate}[I.]
	\item $M_{\mathrm{g}}(<s)=f_gM_{\mathrm{DM}}(<s)$ 
	\item $\rho_{\mathrm{g}}(s^{-})=\Gamma_\mathrm{g}\rho_{\mathrm{g}}(s^{+})$ 
	\item $\rho_{\mathrm{g}}(r)=f_g\rho_{\mathrm{DM}}(r)$ for $r>s$, and for all $r$ if $\Gamma_{\mathrm{g}}=1$  
\end{enumerate}
where the minus and plus indicate the regions just to the inside and outside of the shock, respectively, and we assume that a dark matter jump is coincident with the gas shock, so that $\Gamma_{\mathrm{DM}}=\rho_{\mathrm{DM}}(s^{-})/\rho_{\mathrm{DM}}(s^{+})$; hereafter, for simplicity, we will refer to the dark matter profile with this jump by the name of the unperturbed dark matter model, so we note that, for a given dark matter model $\rho_{\mathrm{m}}(r)$ internal to $s$, the profile external to $s$ is in fact $\rho_{\mathrm{DM}}(r)=\rho_{\mathrm{m}}(r)/\Gamma_{\mathrm{DM}}$. 

Lastly, to allow us to solve for a profile, we introduce one additional assumption: denoting $x=r/r_s$, where $r_s$ is a scale radius in a given model, and referring to condition I, we employ the following ansatz:
\begin{align}\label{e:ansatz}
\frac{1}{f_g}M_{\mathrm{g}}(x)=M_{\mathrm{DM}}({\xi}x^n),
\end{align}
where $\xi$ and $n$ are parameters to be found. This choice of parametrization has the benefits of being mathematically simple and preserving scale invariance. We now apply these conditions to find profiles for the gas internal to the virial shock. 

\subsection{General Derivation}
Beginning with the relation defined by equation (\ref{e:ansatz}), we obtain the gas density profile as:
\begin{align}
\rho_{\mathrm{g}}(x)&=\frac{1}{4{\pi}x^2}\frac{dM_{\mathrm{g}}}{dx}=\frac{f_g}{4{\pi}x^2}\frac{d}{dx}M_{\mathrm{DM}}({\xi}x^n),\\
\rho_{\mathrm{g}}(x)&=f_gn\xi^{3}x^{3n-3}\rho_{\mathrm{DM}}({\xi}x^n).
\end{align}
Now, using conditions II and III, we see that:
\begin{align}
\rho_{\mathrm{g}}\left(\frac{s}{r_s}\right)&=f_gn\xi^{3} \left[\frac{s}{r_s}\right]^{3n-3}\rho_{\mathrm{DM}}\left(\xi\left[\frac{s}{r_s}\right]^n\right);
\end{align}
the argument of the dark matter function then yields:
\begin{align}
\xi\left[\frac{s}{r_s}\right]^n=\frac{s}{r_s}\Rightarrow\xi=\left[\frac{s}{r_s}\right]^{1-n},
\end{align}
and the normalization gives:
\begin{align}
n\xi^{3}\left[\frac{s}{r_s}\right]^{3n-3}=\frac{\Gamma_{\mathrm{g}}}{\Gamma_{\mathrm{DM}}}\equiv\Gamma.
\end{align}
Combining these two expressions then provides us with $\xi$ and $n$ in terms of our physical variables:
\begin{align}
n&=\Gamma,\\
\xi&=\left(\frac{s}{r_s}\right)^{1-\Gamma}.
\end{align}
This means that our final expression for the gas density is simply:
\begin{align}\label{e:general_exp}
\rho_{\mathrm{g}}\left(\frac{r}{r_s}\right)&={\Gamma}f_g\left(\frac{r}{s}\right)^{3\Gamma-3}\rho_{\mathrm{DM}}\left(\frac{s}{r_s}\left[\frac{r}{s}\right]^{\Gamma}\right).
\end{align}

\subsection{Power Law Profile}
To motivate our method, we first obtain a family of gas density profiles assuming a simple power law for the dark matter density profile,
\begin{align}
\rho_{\mathrm{DM}}(r)=\frac{A}{r^k},
\end{align}
with $k<3$. This profile describes, for instance, the singular isothermal sphere with $k=2$. For this profile, $r_s=1$, and equation (\ref{e:general_exp}) yields:
\begin{align}
\rho_{\mathrm{g}}(r)=\frac{{\Gamma}f_gA}{s^k(r/s)^{3+\Gamma(k-3)}}.
\end{align}
It is straightforward to verify that such a profile fulfills the conditions of $\S$\ref{s:assumptions}.

\subsection{Einasto Profile}
We next consider the Einasto profile \citep{einasto65}, 
\begin{align}
\rho_{\mathrm{DM}}(r)=A\exp\left[-\frac{2}{\alpha}\left(\frac{r}{r_s}\right)^{\alpha}\right],
\end{align}
which recent cosmological simulations have found to be a very good description of dark matter halos \citep[e.g.,][]{ludlow13}, but which has not yet enjoyed as widespread use as some other models, partly due to the lack of closed-form expressions for many quantities of interest, such as the surface density \citep{retana12}. However, the gas density derived from this profile using our method does have an analytical form:
\begin{align}
\rho_{\mathrm{g}}(r)={\Gamma}f_gA\left(\frac{r}{s}\right)^{3\Gamma-3}\exp\left[-\frac{2}{\alpha}\left(\frac{s}{r_s}\right)^{\alpha}\left(\frac{r}{s}\right)^{\Gamma\alpha}\right].
\end{align}

\subsection{NFW Profile}\label{s:nfw}
We now apply our method to the widely-used Navarro-Frenk-White (NFW) \citep{nfw} profile,
\begin{align}
\rho_{\mathrm{DM}}(r)=\frac{A}{r/r_s(1+r/r_s)^2},
\end{align}
which has been demonstrated to provide a good fit to dark matter halos in simulations. Using equation (\ref{e:general_exp}), we find:
\begin{align}\label{eq:GammaModel}
\rho_{\mathrm{g}}(r)={\Gamma}f_gA\frac{\left(r/s\right)^{2\Gamma-2}}{r/r_s\left(1+\left(s/r_s\right)\left(r/s\right)^{\Gamma}\right)^2}.
\end{align}

\begin{figure}
\figurenum{1}
\epsscale{1.2}
\plotone{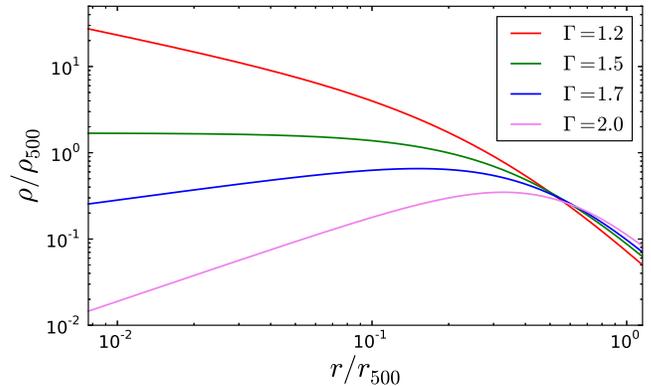}\label{f:cluster_models}
\caption{Gas profiles for various choices of $\Gamma$, assuming an NFW dark matter profile as described in $\S$\ref{s:nfw}.}
\end{figure}

Our profile has two parameters fixed by the choice of an NFW dark matter profile, the amplitude $A=\rho_c\delta_c$ and the scale radius $r_s$. These can be obtained independently of the gas data (for instance, via weak lensing). The other three are free parameters: the gas fraction $f_g$, the shock radius $s$, and the jump ratio $\Gamma$. Qualitatively, our analytic expression for the gas density yields a cuspy inner profile for $1<\Gamma<1.5$, a pure core for $\Gamma=1.5$, a core with an inner decline for $1.5<\Gamma\lesssim{2}$, and a central depression for $\Gamma\gtrsim{2}$. Examples of these profile types are illustrated in Figure~\ref{f:cluster_models}.

\subsection{Generalized NFW Profiles}\label{s:gNFW}
The NFW profile of the preceding section can be characterized by its behavior at two extremes -- in the large $r$ limit, $\rho_{\mathrm{NFW}}\sim{r^{-3}}$, while for small $r$, $\rho_{\mathrm{NFW}}\sim{r^{-1}}$. However, dark matter halos do exhibit some scatter, particularly in their internal structure. As a result, it is possible to consider the NFW profile as a special case of a broader family of models, the generalized NFW profiles:
\begin{align}
\rho(r)=\frac{A}{\left(r/r_s\right)^{\alpha}\left(1+r/r_s\right)^{3-\alpha}},
\end{align}
which ensure that for large $r$, $\rho\sim{r^{-3}}$, while allowing for variations in the central region, for which $\rho\sim{r^{-\alpha}}$.

Both cosmological simulations and observational data have yielded diverse estimates for the value of the $\alpha$ parameter. As discussed above, the work of \cite{nfw} yielded $\alpha=1$, in good agreement with simulations and data. On the other hand, the simulations of \cite{moore99} suggested that $\alpha=1.5$, which is an upper bound of the range of the recent results of \cite{schaller14}, whose simulations found dark matter slopes that tend to be steeper than that of the NFW model for small clusters, and similar to NFW for large clusters. On the observational side, several efforts have obtained values of $\alpha<1$ \citep[][and references therein]{newman13}, while other analyses have concluded that $\alpha\approx0.9-1.2$ \citep{oguri09,richard09,saha09}.

We can test the implications of these variations for our method by deriving a family of profiles for the generalized NFW profile. From equation (\ref{e:general_exp}), we obtain:
\begin{align}\label{e:gNFW_gas}
\rho_{\mathrm{g}}(r)={\Gamma}f_gA\frac{(r/s)^{(3-\alpha)\Gamma+\alpha-3}}{(r/r_s)^{\alpha}(1+(s/r_s)(r/s)^{\Gamma})^{3-\alpha}}.
\end{align}
We immediately see that this profile reduces to equation (\ref{eq:GammaModel}) for $\alpha=1$. 

In principle, we are interested in the regime in which the profiles yield either cuspy or cored interiors, which are the two observed types of gas density profile, as will be explored further in $\S$\ref{s:obscomp}. From equation (\ref{e:gNFW_gas}), this condition can be expressed as:
\begin{align}
(3-\alpha)\Gamma-3\lesssim0,
\end{align}
which imposes a constraint on the allowed values of $\Gamma$. We thus find that, for $\alpha=1$, as before, $\Gamma\lesssim{3/2}$, while for $\alpha=3/2$, $\Gamma\lesssim{2}$, and for $\alpha=2$, $\Gamma\lesssim{3}$. Accordingly, we see that by modifying the value of $\alpha$, we can alter the range of acceptable $\Gamma$ values.  However, for the remainder of the paper, we will use the gas profile that we obtained from the standard NFW ($\alpha=1$) dark matter profile, defined by equation (\ref{eq:GammaModel}).

\section{Comparison with Observations}\label{s:obscomp}
Having established the mathematical structure of our model (equation (\ref{eq:GammaModel})), we now aim to test it against observational data. We use deprojected gas density profiles from \cite{av2006} (hereafter V06), which consists of 13 low-redshift, relaxed clusters and groups, and the low-redshift sample of \cite{av2009} (V09), which provides a more representative sample of clusters, including both relaxed and disturbed systems. We defer to the above papers for additional details on the data, but we note that we adopt the same $\Lambda$CDM cosmological parameters: $h=0.72$, $\Omega_{\Lambda}=0.7$, and $\Omega_{m}=0.3$.

We fit both our model and the $\beta$-model to the data, excluding USGC S152 since it has no listed $r_{500}$ -- the radius corresponding to a density of 500 times the critical density $\rho_c$ -- or $c_{500}$ -- the corresponding concentration parameter -- value. For the clusters in the 2006 sample, we have all of the quantities -- the redshift $z$, $c_{500}$, and $r_{500}$ -- needed to determine the NFW parameters of our model. For the V09 sample, we lack only individual $c_{500}$ values, and instead assume a fixed value of $c_{500}=3.5$ for all the clusters, which is reasonable for the mass range of these clusters (see V06 and V09 and references therein). 

\begin{figure*}
\figurenum{2}
\begin{center}
\includegraphics[scale=0.29]{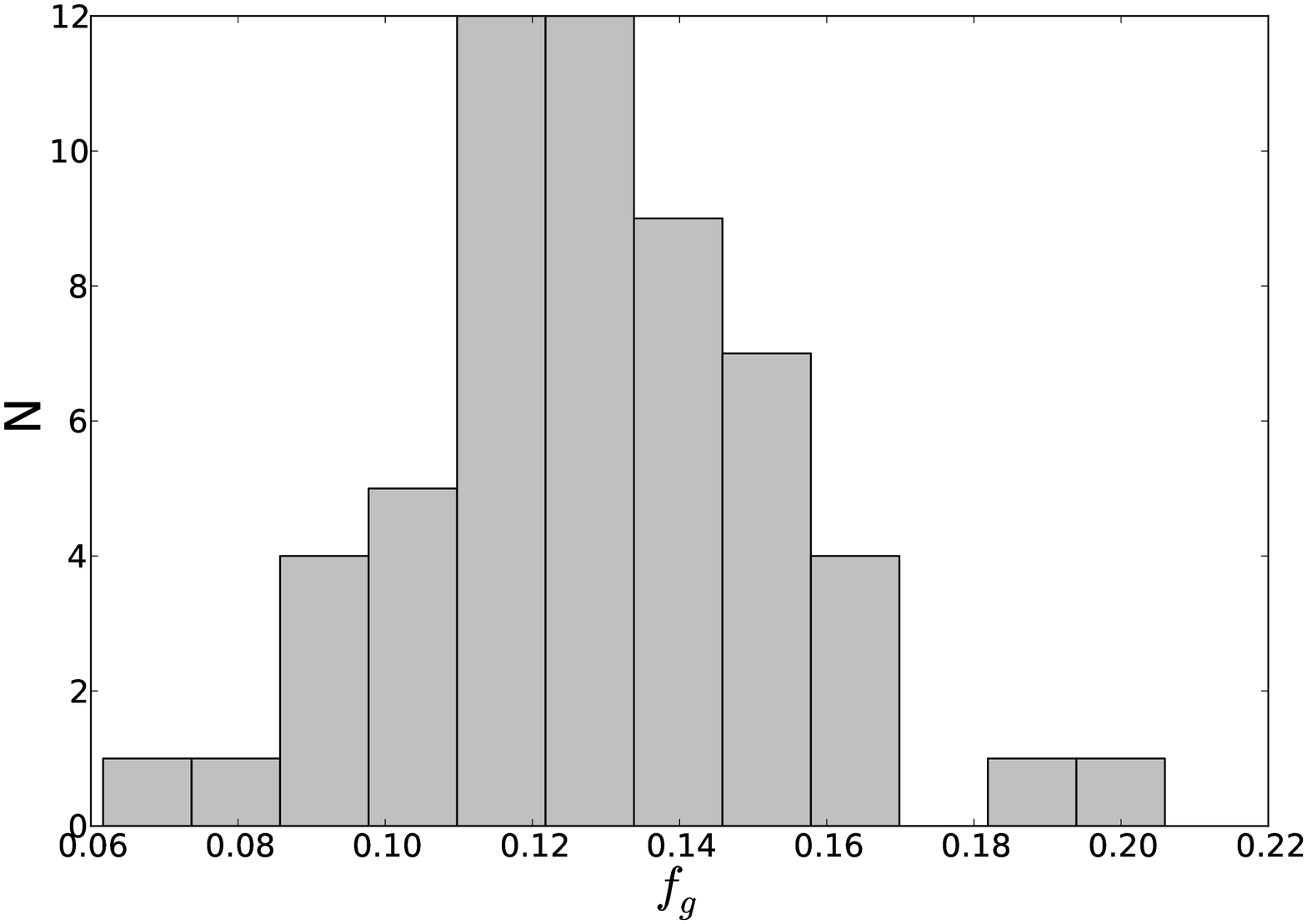}
\includegraphics[scale=0.29]{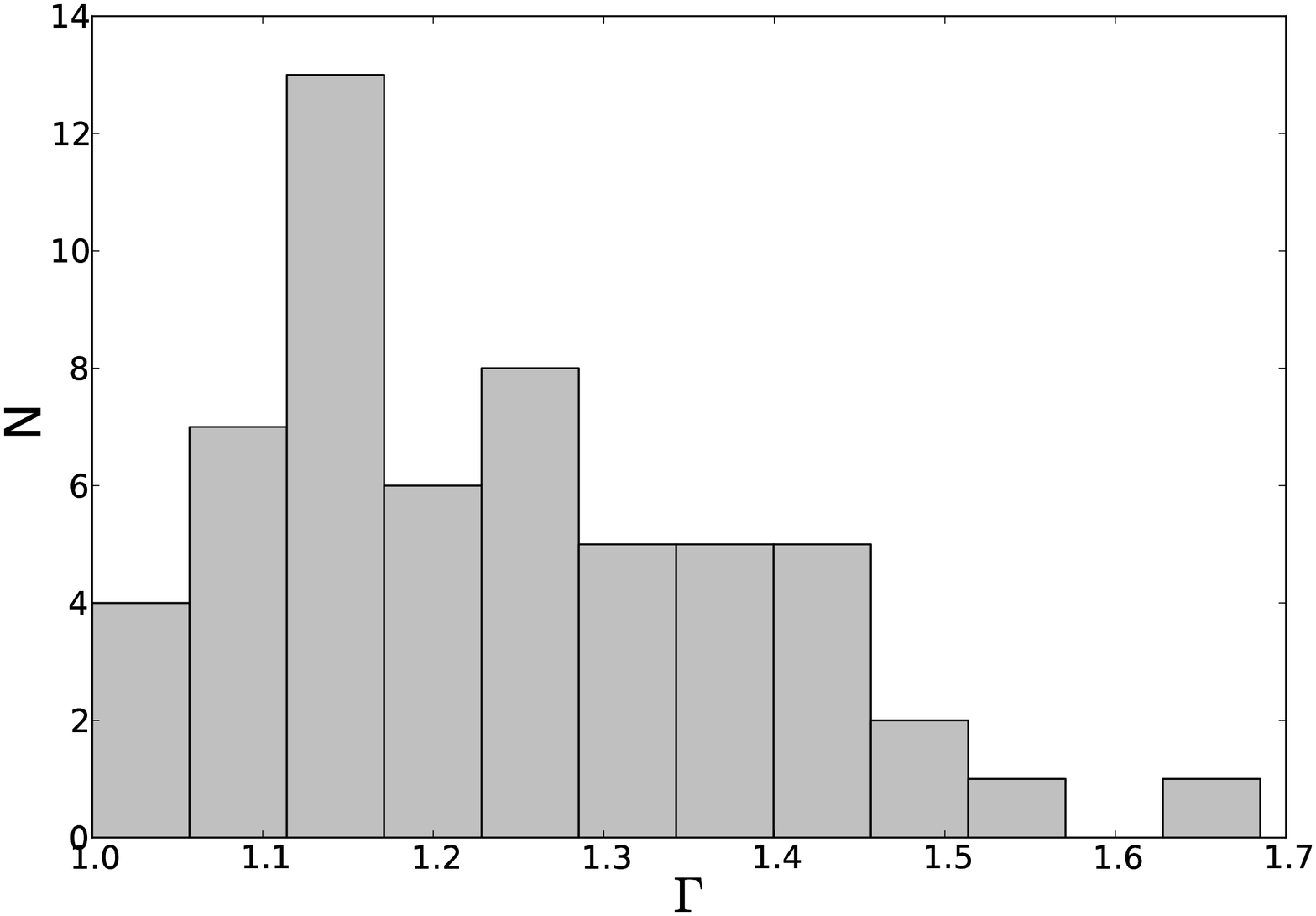}\hspace{0.1in}
\includegraphics[scale=0.32]{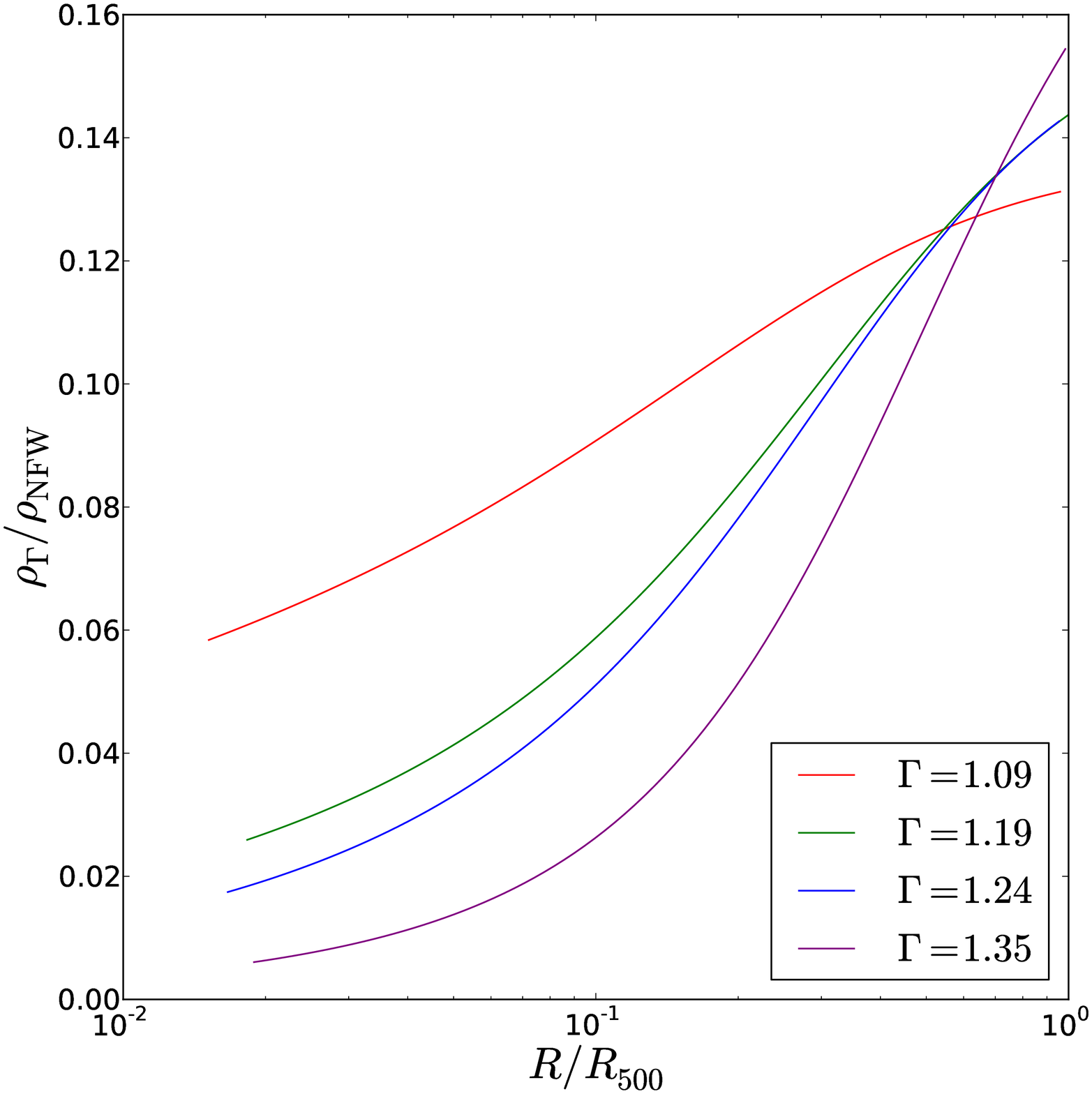} 
\includegraphics[scale=0.35]{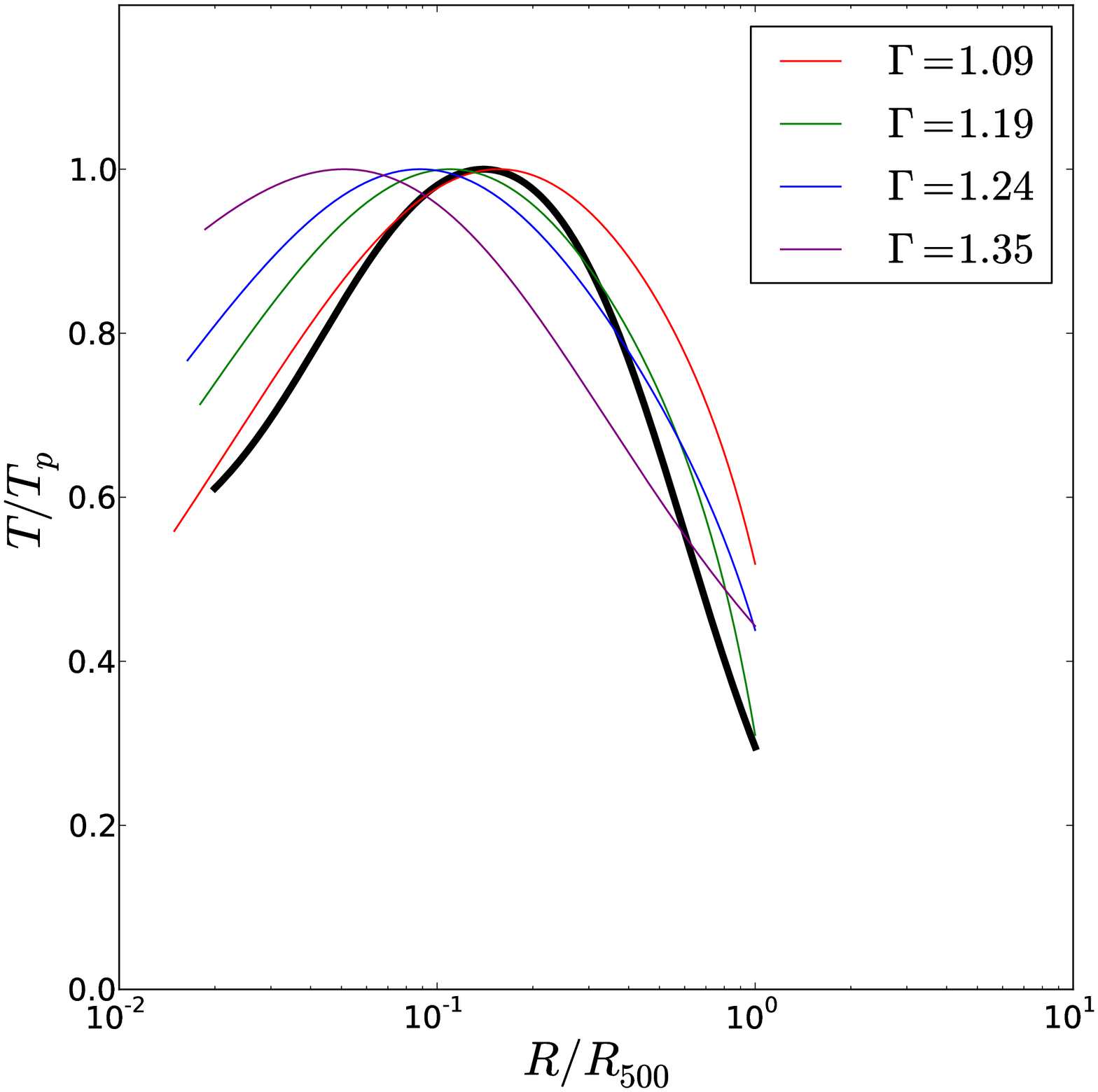}\vspace{-0.1in}
\end{center}
\caption{A summary of the results of fitting our model to the cluster sample. Top: the distribution of fitted $f_g$ (left) and $\Gamma$ values (right) for the V06 and V09 samples. The parameter $s$ is poorly constrained and thus not included. Bottom: the radial variation of $\rho_{\Gamma}/\rho_{\mathrm{NFW}}$ (left) and the scaled temperature profiles (right) resulting from a numerical computation using the equation of hydrostatic equilibrium and our density model for four clusters alongside the average temperature profile from V06 in black.}\label{f:fitpars}
\end{figure*}

\begin{figure*}
\figurenum{3}
\epsscale{0.41}
\plotone{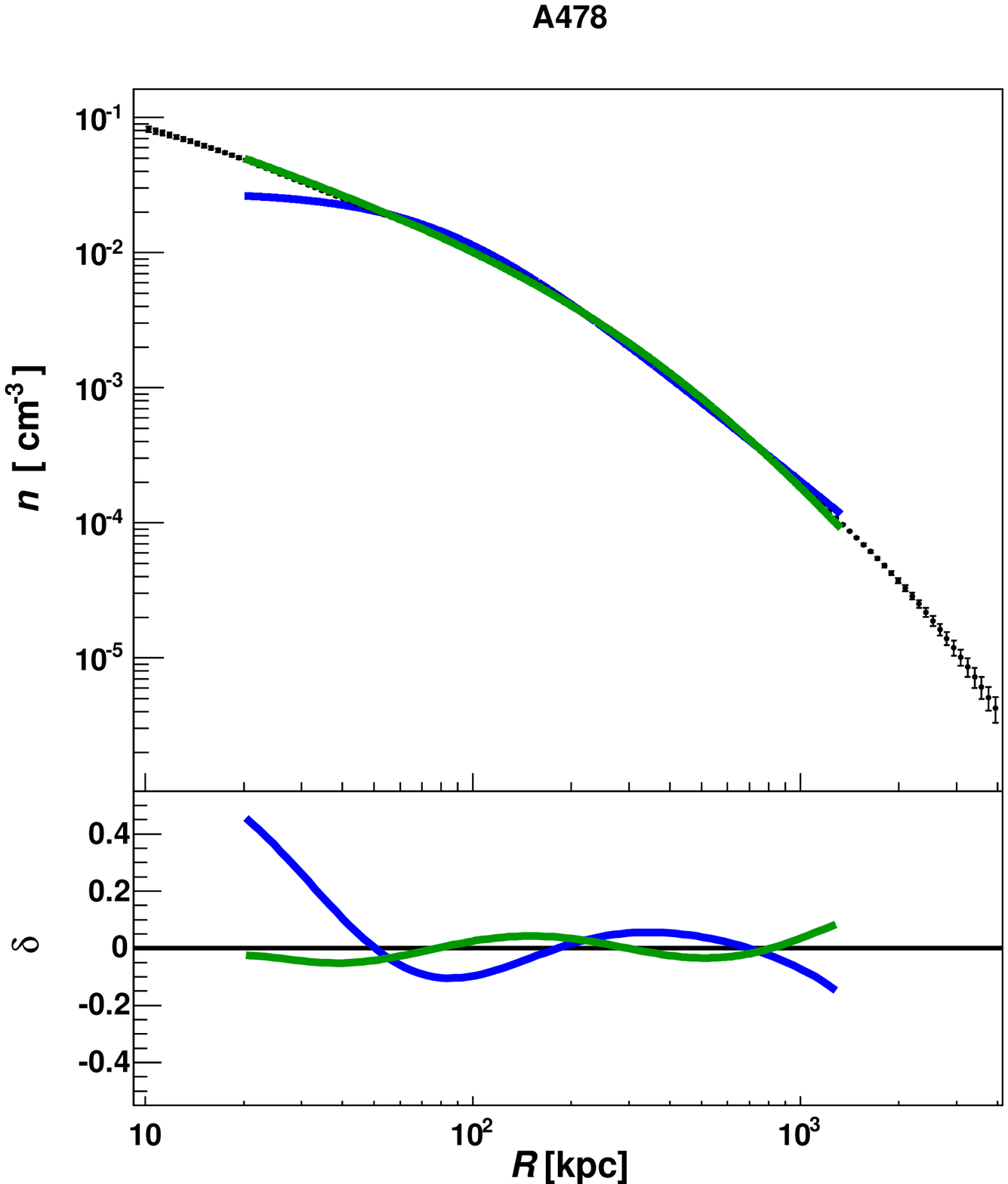}\hspace{-0.3in} 
\plotone{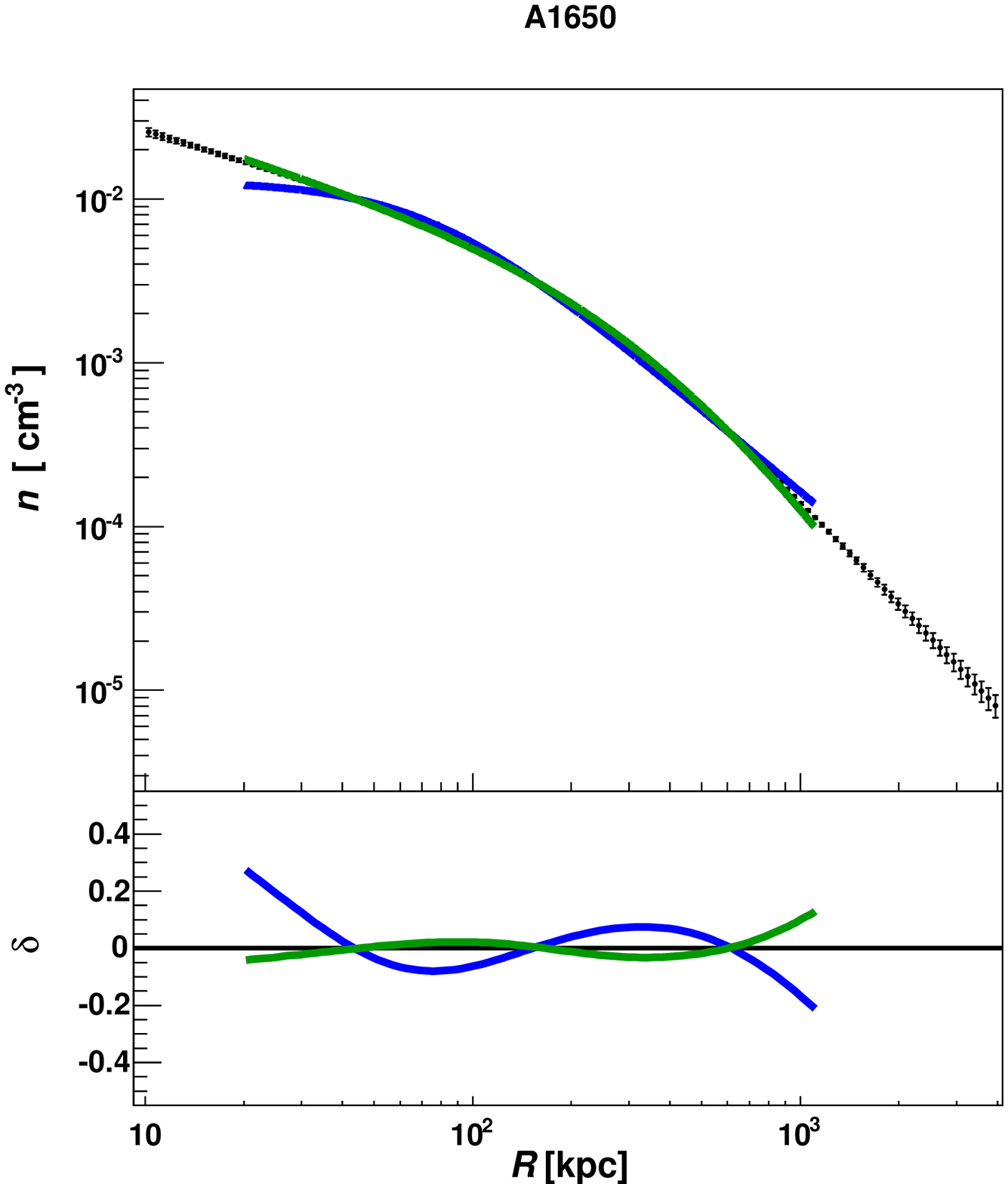}\hspace{-0.3in} 
\plotone{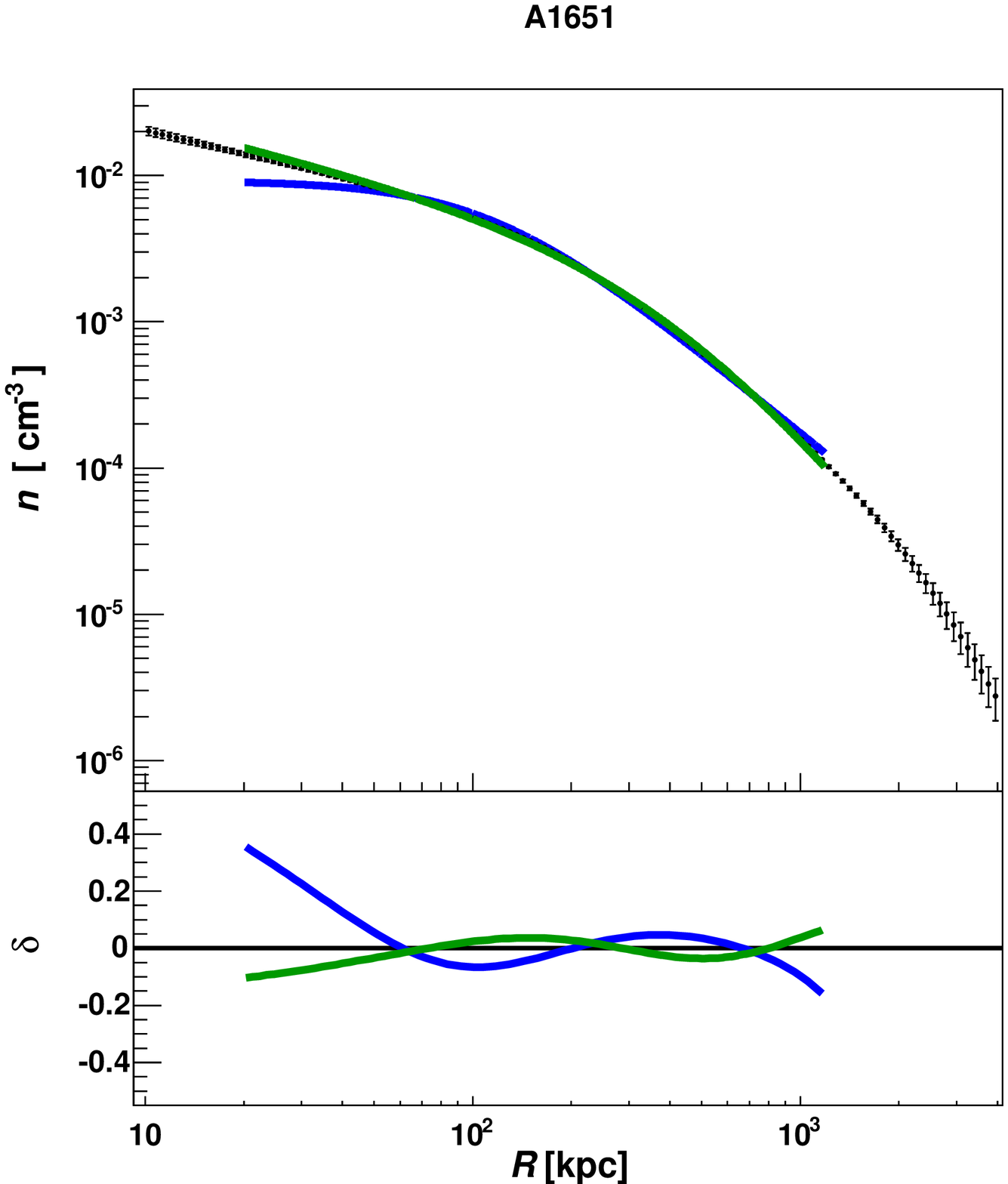}\hspace{-0.3in} 
\plotone{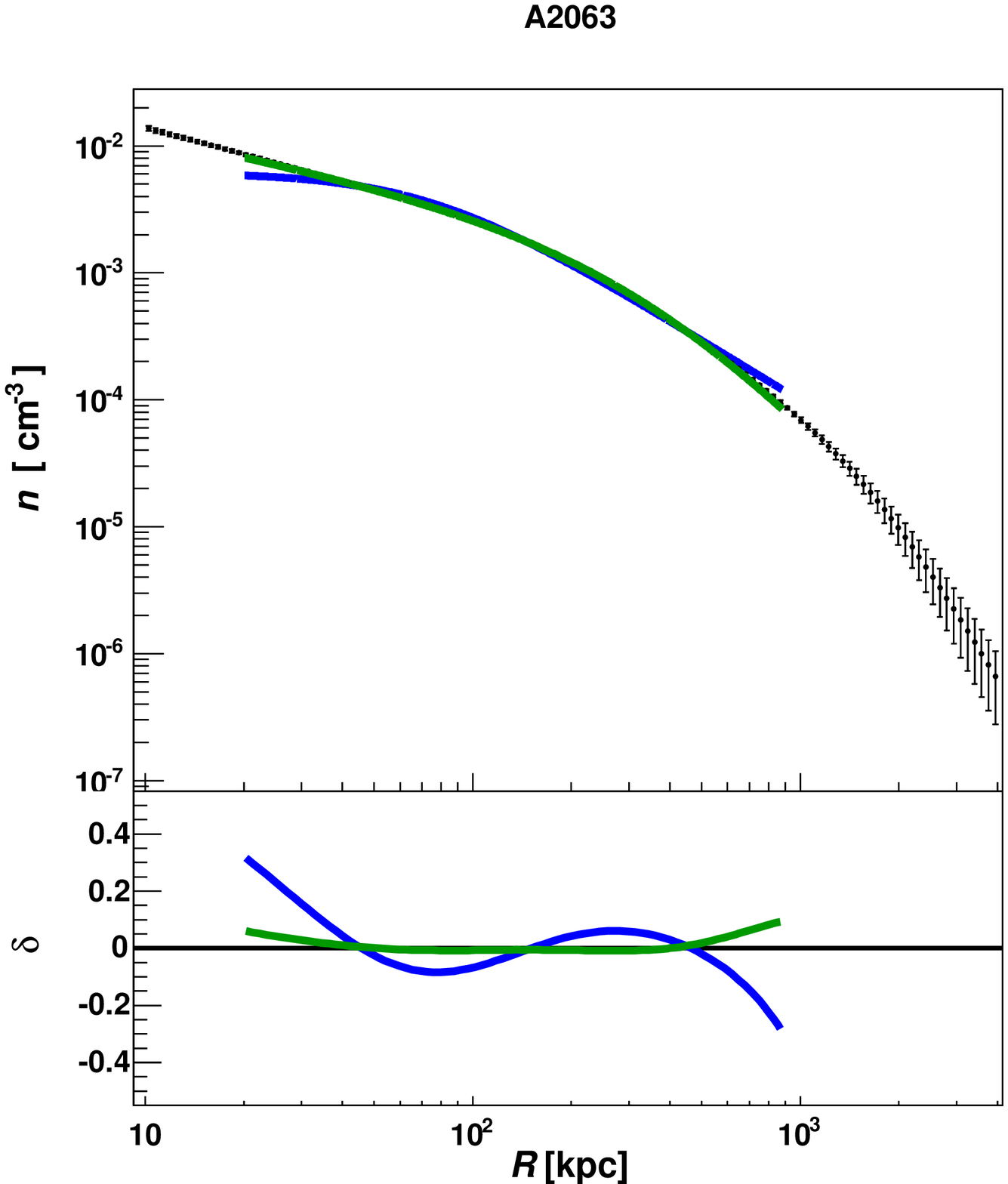}\hspace{-0.3in} 
\plotone{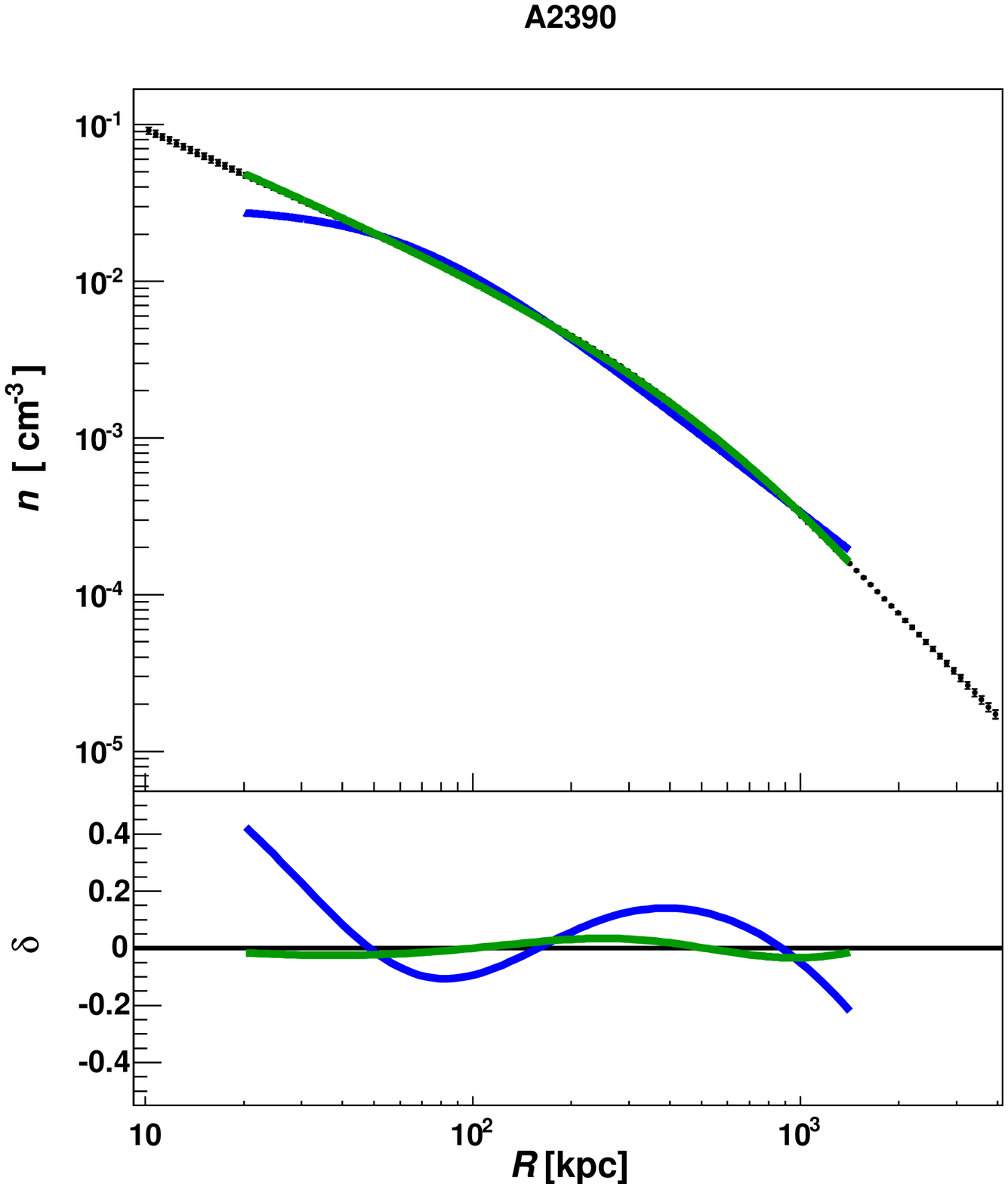}\hspace{-0.3in} 
\plotone{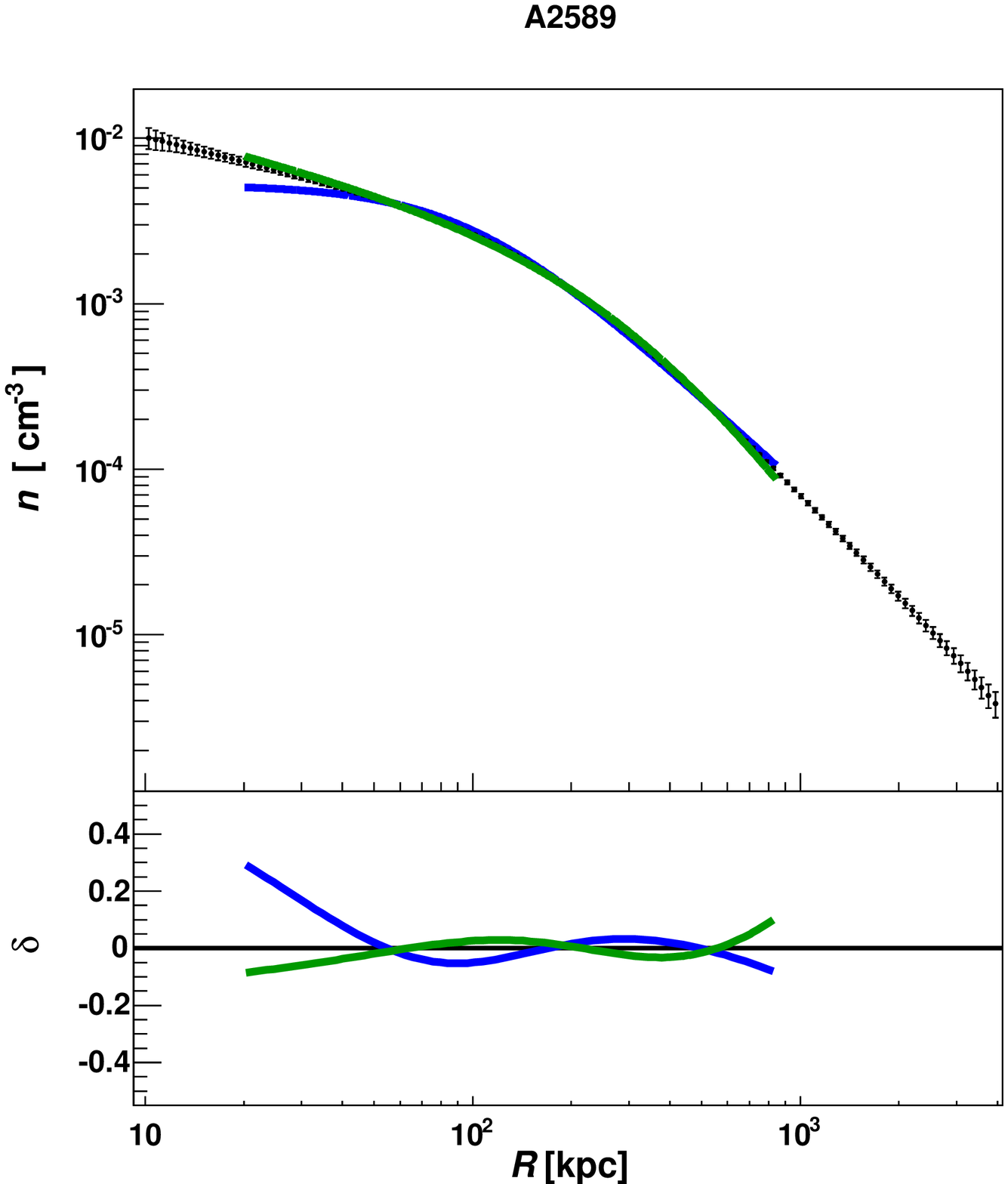}\hspace{-0.3in} 
\plotone{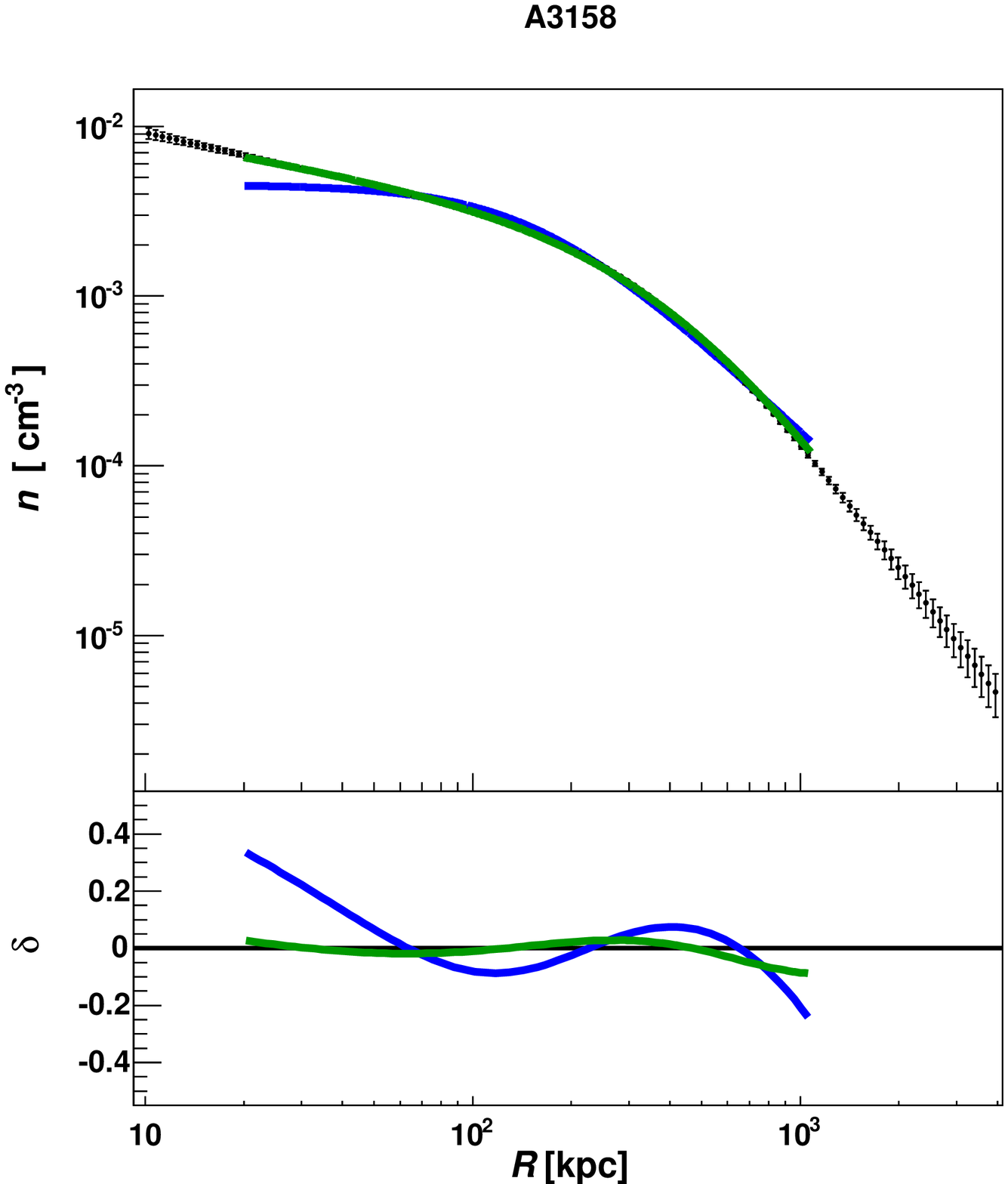}\hspace{-0.3in} 
\plotone{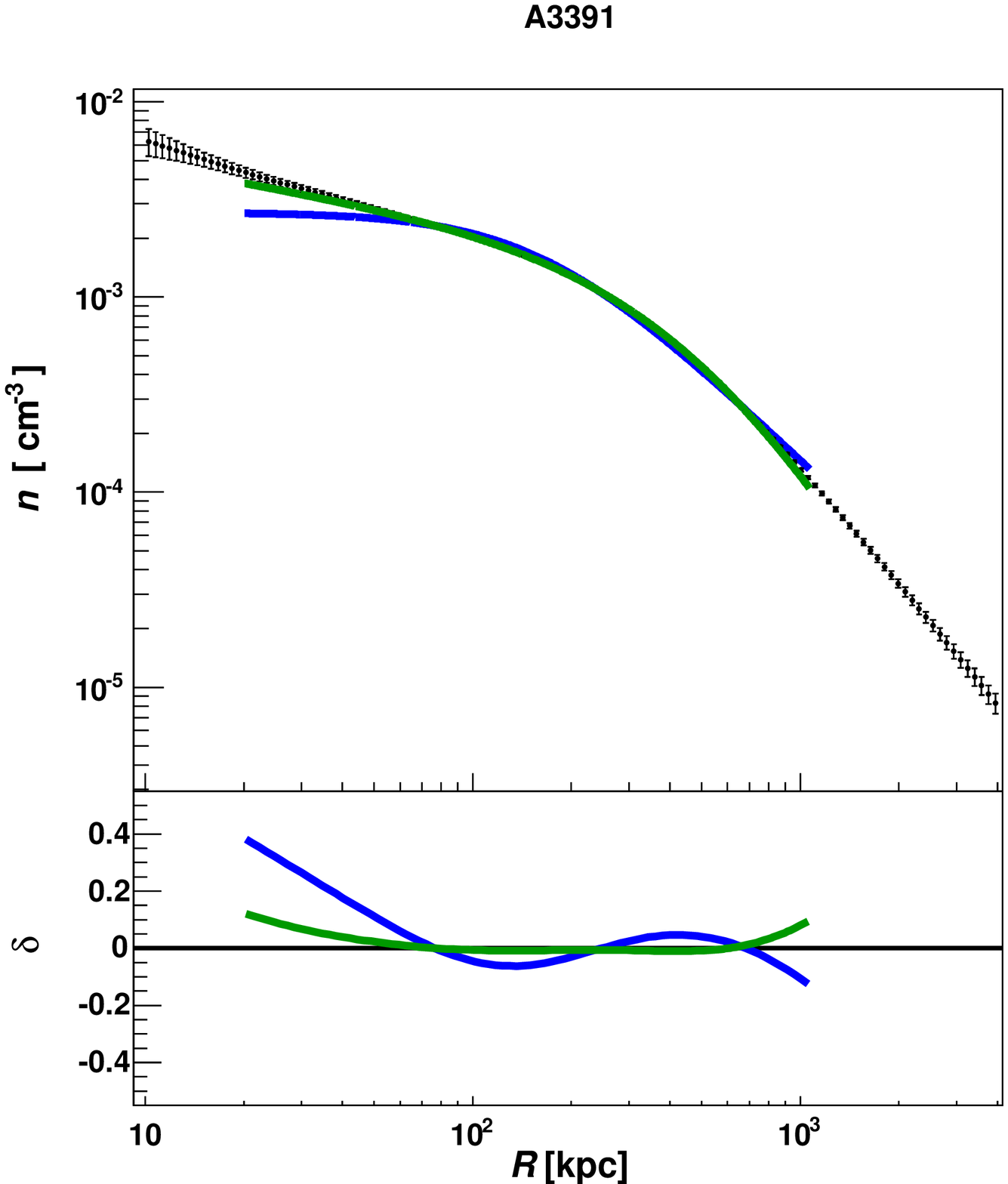}\hspace{-0.3in} 
\plotone{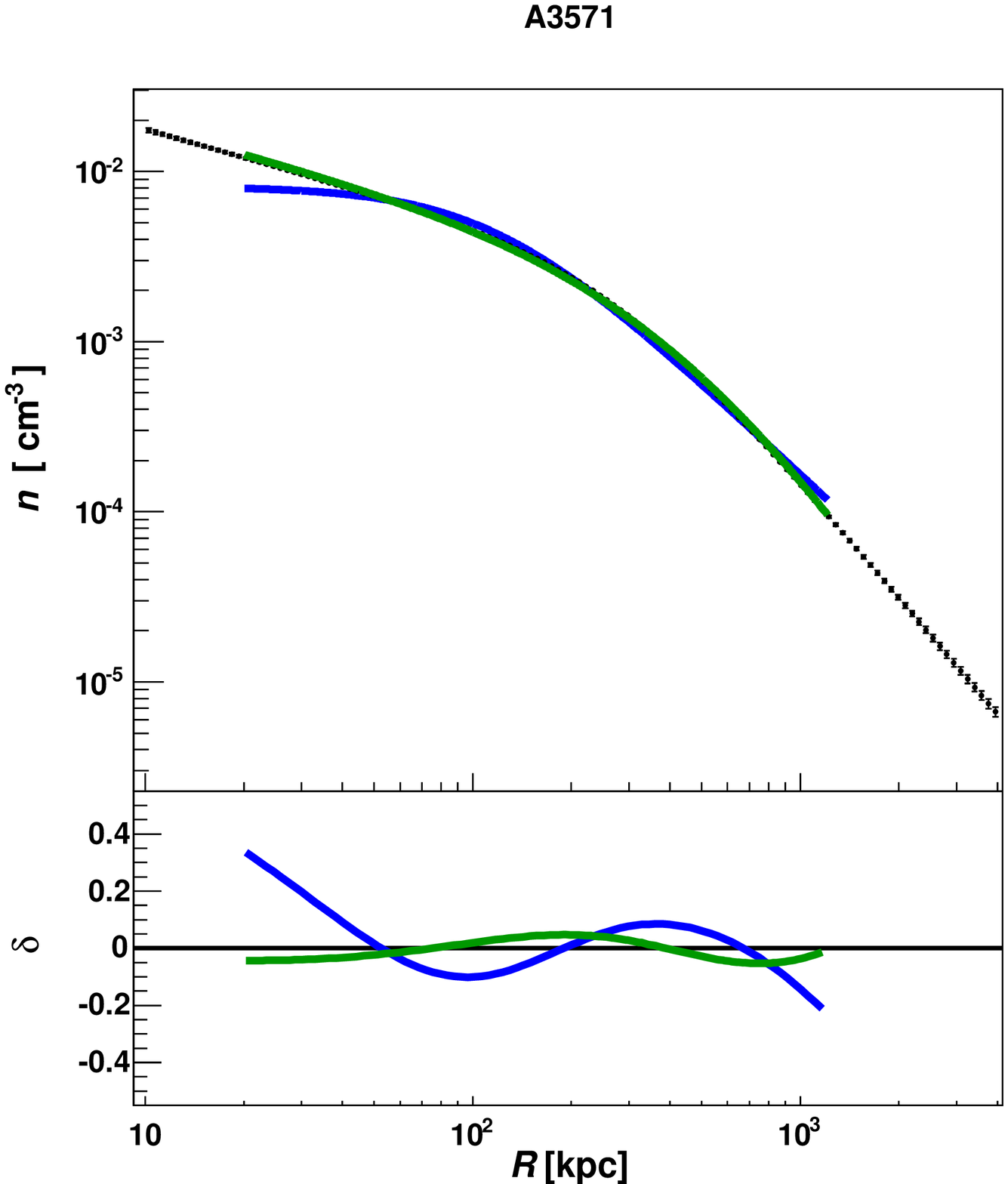}
\caption{Examples of fits to a subsample of gas density profiles of relaxed clusters for which our model (green) provides a reasonable description over the range of radii considered; the $\beta$-model (blue) is shown for comparison. The lower panel shows the normalized residual, $\delta$ (see text).}\label{f:rel_clusters}
\end{figure*}

\begin{figure*}
\figurenum{4}
\epsscale{0.41}
\plotone{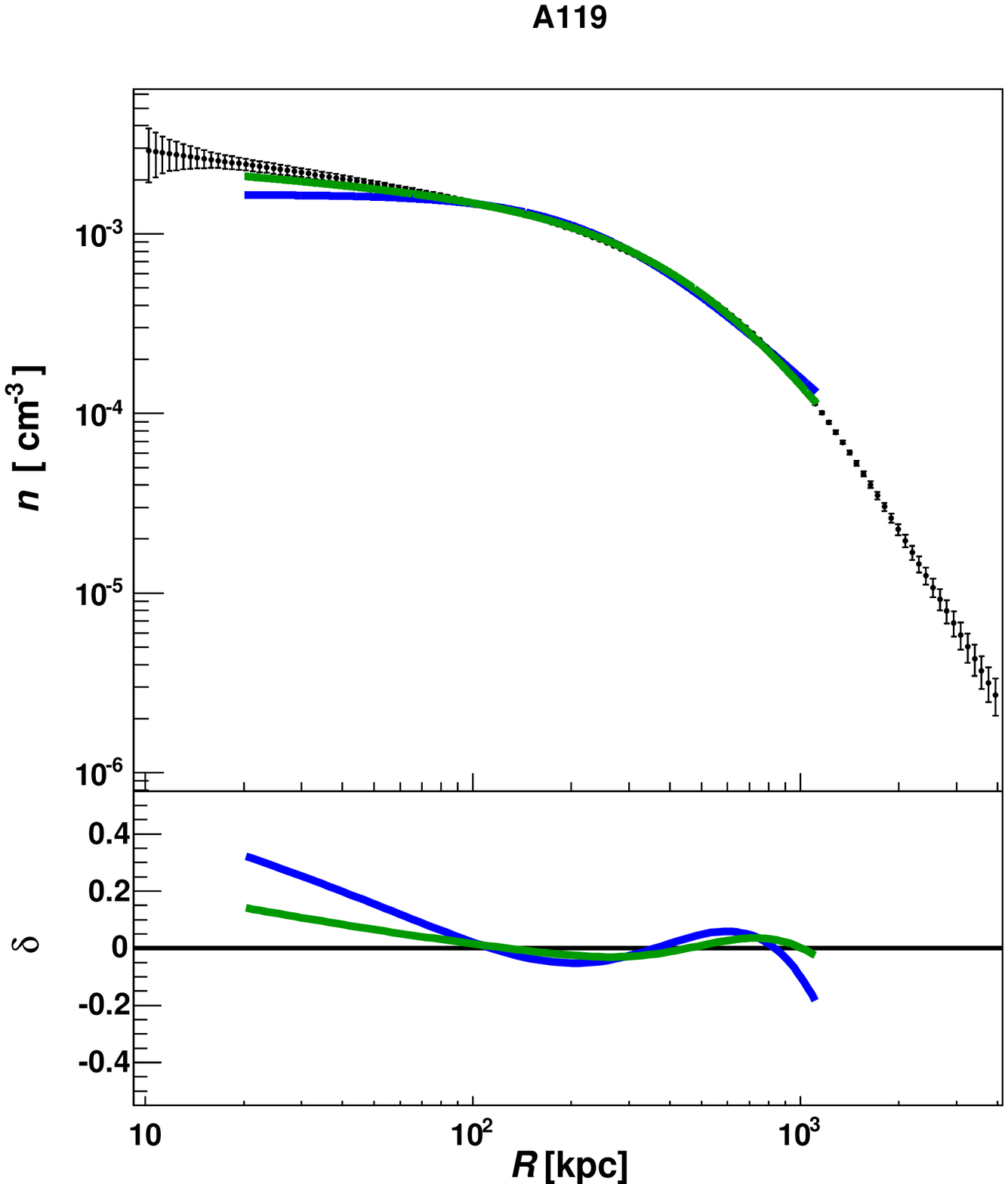}\hspace{-0.3in}
\plotone{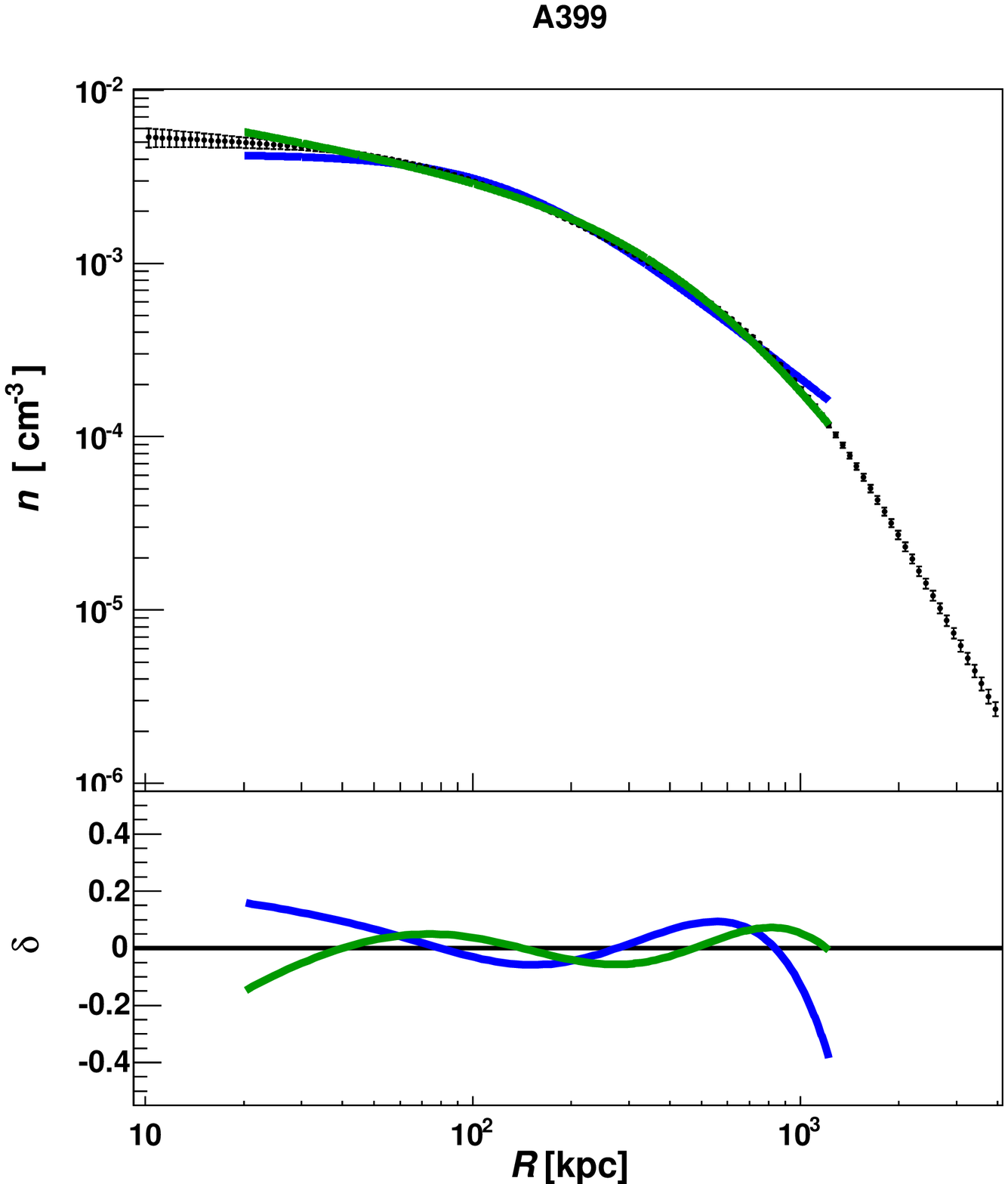}\hspace{-0.3in}
\plotone{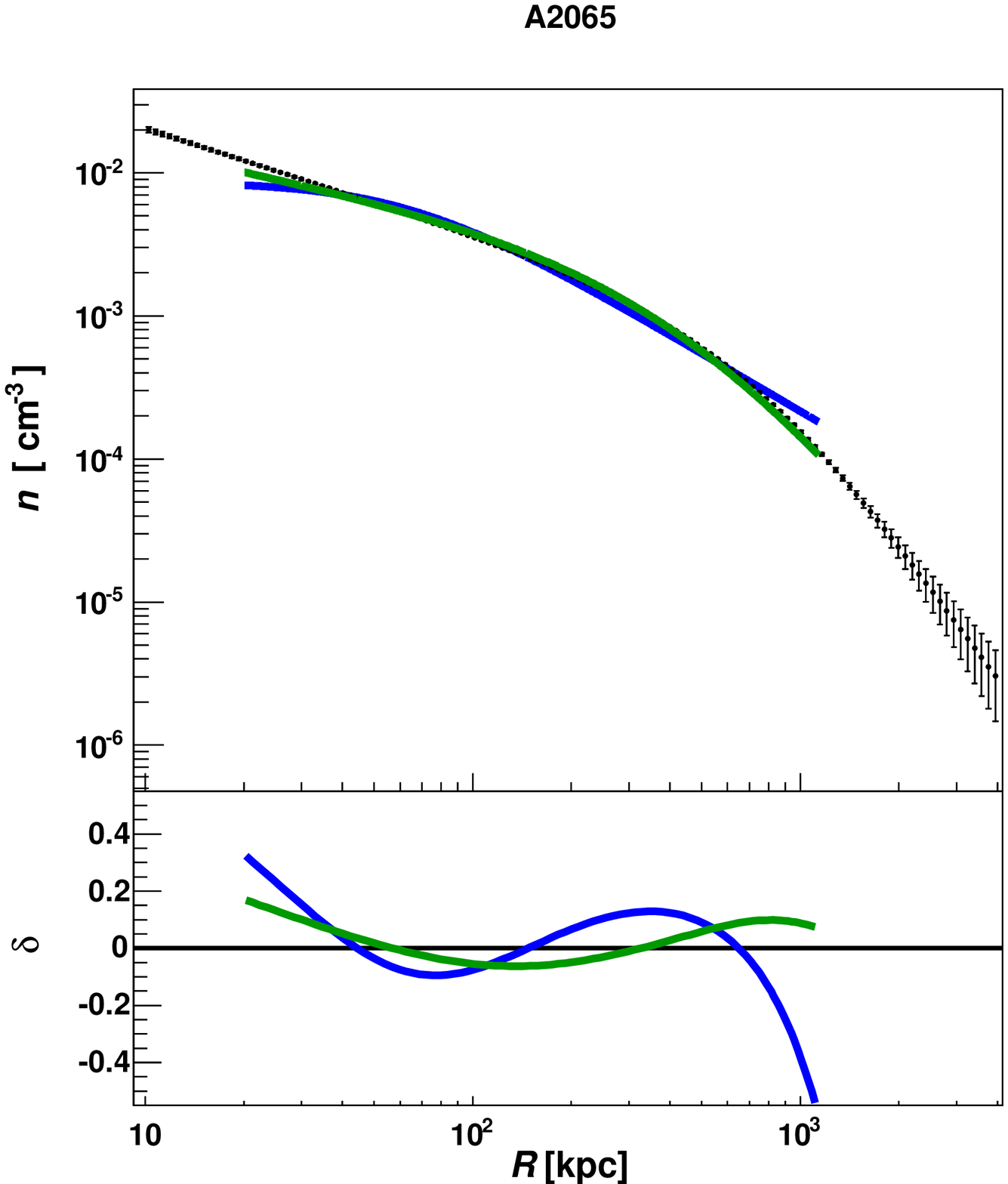}\hspace{-0.3in}
\plotone{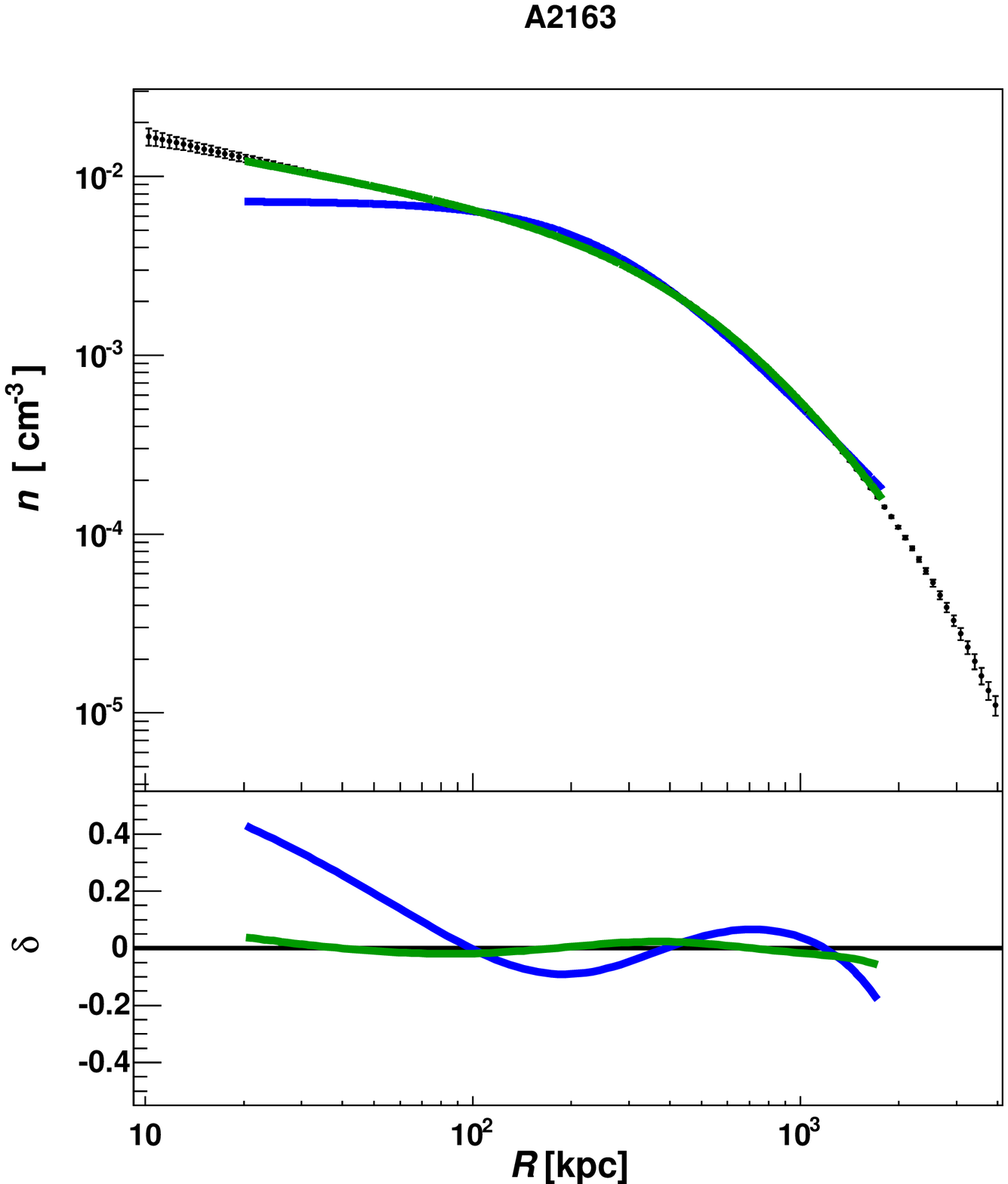}\hspace{-0.3in}
\plotone{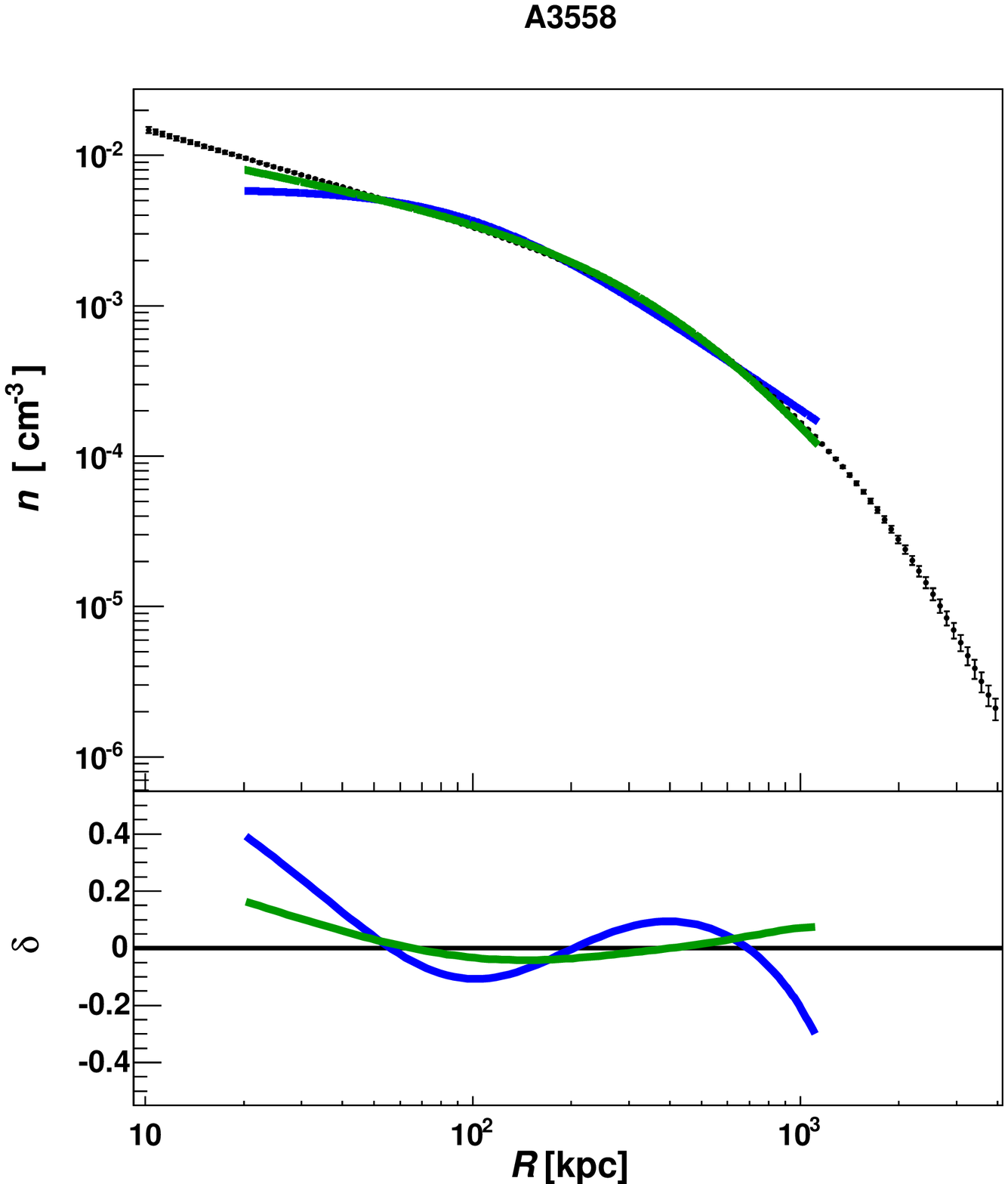}\hspace{-0.3in}
\plotone{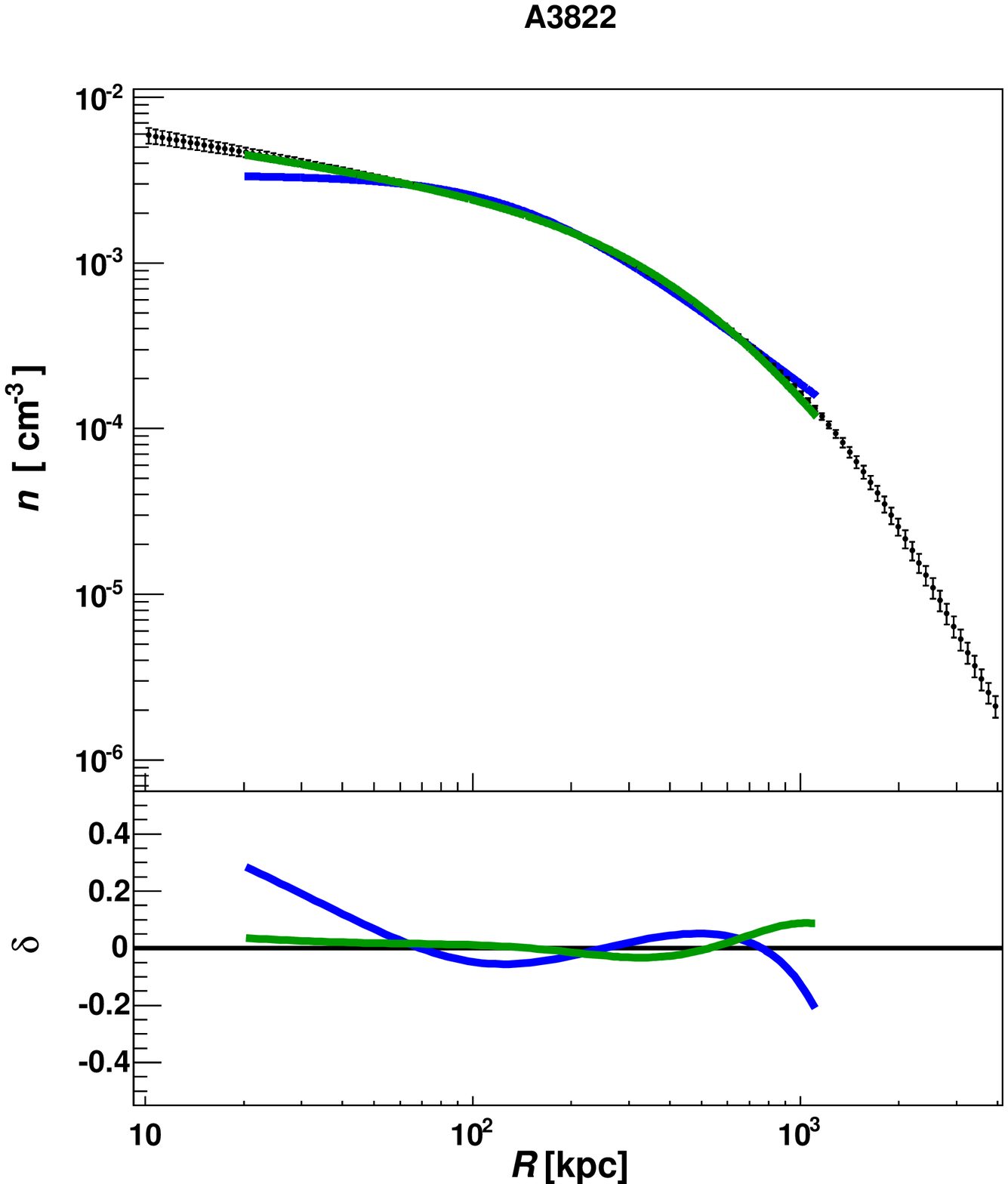}
\caption{Same as Figure~\ref{f:rel_clusters}, but for systems identified as mergers by V09.}\label{f:merging_clusters}\vspace{0.2in}
\end{figure*}

To transform the mass density to a number density, we use $\rho_g=1.274m_p(n_en_p)^{1/2}$, where the conversion factor assumes a primordial abundance of helium and a $Z=0.1Z_{\odot}$ abundance of heavier elements. We additionally constrain our fits to $20\;\mathrm{kpc}<r<r_{500}$; the lower limit is imposed to avoid the contribution of the central galaxy, and the upper limit excludes the data points that are beyond the field-of-view of \textit{Chandra} and are instead extrapolated from the model of V06. The results of our fits to the clusters are summarized in Figure \ref{f:fitpars} and examples of our fits are shown in Figure~\ref{f:rel_clusters} for relaxed clusters and Figure~\ref{f:merging_clusters} for merging systems. We now discuss these in turn.

\vspace{-0.1in}

 \subsection{Fit Parameters}
The top left panel of Figure~\ref{f:fitpars} shows the distribution of fitted $f_g$ parameters, which are generally reasonable gas fraction estimates. However, it is worth noting that the $f_g$ parameter measured here is the gas fraction interior to the shock radius $s$, as per condition I, as well as the local gas fraction normalization of the profile exterior to $s$, following condition III. Interior to $s$, the gas fraction is not constant (except in the case where $\Gamma_\mathrm{g}=1$ and the interior profile is simply a scaled NFW profile). The radial variation of the ratio of our profile to its progenitor NFW profile is shown in the bottom left panel of Figure~\ref{f:fitpars} for four clusters with various values of $\Gamma$.

The distribution of fitted $\Gamma$ parameters is given in the top right panel of Figure~\ref{f:fitpars}, from which it is clear that the data favor values of $\Gamma\lesssim{1.5}$. This trend can be partly explained by the composition of our sample, which comprises primarily relaxed clusters with cuspy central profiles. This suggests that $\Gamma_\mathrm{g}$ and $\Gamma_\mathrm{DM}$ should be comparable. The recent simulations of~\cite{schaal14} favor ${\cal{M}}\approx2.7$, which for a gas with $\gamma=5/3$ implies a shock jump of $\Gamma_{\mathrm{g}}\approx3$; these simulations coupled with our inferences from X-ray data thus suggest that $\Gamma_{\mathrm{DM}}\approx2$-$3$.
 
An additional consequence of these low $\Gamma$ values is that the model is not very sensitive to the value of $s$; indeed, by construction, if $\Gamma =1$, the $s$ parameter disappears from the model entirely. Accordingly, the $s$ parameter is not well-constrained, and the fitting procedure typically chooses a value at one of the fitting limits. These results thus indicate that our model is really at most a two-parameter model; the use of our model with even fewer parameters is discussed further in $\S$\ref{s:modeltests}.

We present one other result in Figure~\ref{f:fitpars}: in the bottom right panel, we show the temperature profiles obtained for four relaxed clusters by numerically solving the equation of hydrostatic equilibrium with sensible initial conditions for each cluster. For comparison, we scale all the temperature profiles by their peak value $T_p$ and also plot the average analytical model of V06; we see good agreement between the analytical model and our numerical profiles for relaxed clusters. Since this suggests that our model is in agreement with the detailed prescriptions of V06, we now turn to comparisons with the traditional $\beta$-model. 

\subsection{Model Comparison}
Figures~\ref{f:rel_clusters} (relaxed clusters) and \ref{f:merging_clusters} (merging clusters) show a subset of clusters fitted with both our model and the $\beta$-model for which our model provides a good description of the behavior of the gas over the fairly wide range of radii we consider. The bottom panel in each of the plots shows the normalized residual $\delta=(\rho_m-\rho_d)/\rho_d $, where $m$ and $d$ refer to the model and data, respectively. These residuals provide a measure for comparing the fits, as we cannot use the reduced $\chi^2$, for not only are we fitting to data reconstructed via the analytical model of V06, but we are doing so with a model whose parameters are non-linear, rendering the statistic formally untenable.

\begin{figure*}
\figurenum{5}
\begin{center}
\includegraphics[scale=0.32]{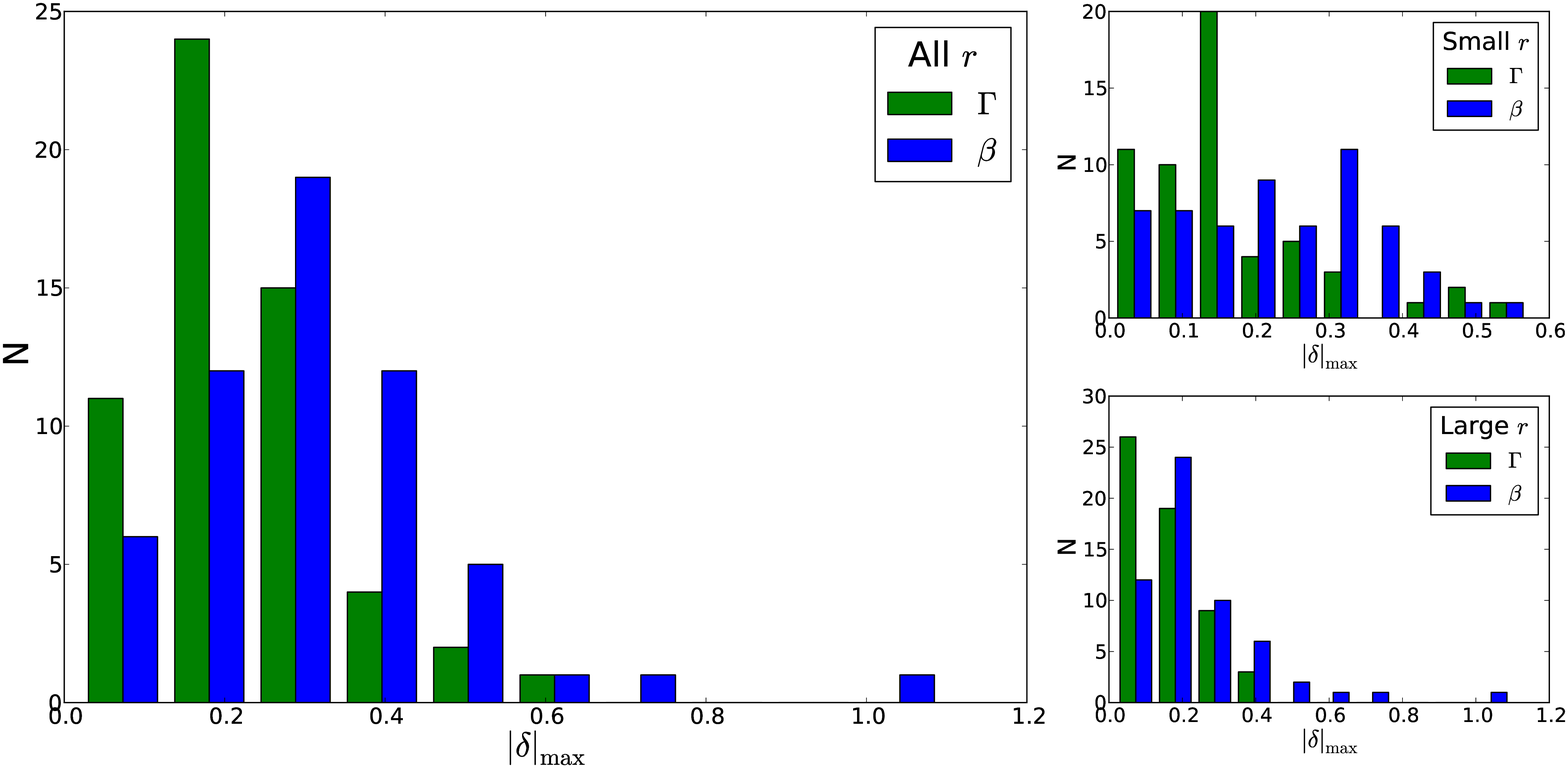}
\end{center}
\caption{Histograms of the maximum normalized residual $|\delta|$ from fitting our model (green) and the $\beta$-model (blue) to all clusters. The lefthand panel considers the values of $|\delta|$ over the entire radial range of each cluster ($20\;\mathrm{kpc}<r<r_{500}$), while the righthand panels show the distributions using truncated ranges: the central region (top), $20\;\mathrm{kpc}<r<{400}\;\mathrm{kpc}$, and the outer region (bottom), $400\;\mathrm{kpc}<r<r_{500}$.}\label{f:residuals}
\end{figure*}

We make such a comparison in Figure~\ref{f:residuals}, in which we show the distribution of maximum normalized residuals $|\delta|$ for all clusters. The distributions are clearly more favorable for our model. If we compare the values of these maximum residuals, we find, as expected, that our model does better in the central regions where the $\beta$-model cannot accommodate a cusp, but we also see significant improvement using our model in the outer regions of the clusters, where the $\beta$-model yields several large outliers. 

Accordingly, we find that our model provides a good description for many of the clusters in our sample, and improves upon the $\beta$-model for most clusters over the range $20\;\mathrm{kpc}<r<r_{500}$, although our model does tend to not be a good description of the lowest mass clusters in our sample ($M_{500}\lesssim{3}\times10^{14}M_{\odot}$). Additionally, as can be seen by way of comparing Figures~\ref{f:rel_clusters} and \ref{f:merging_clusters}, our model tends to be a better description for relaxed systems than merging ones. This is not particularly surprising, as our derivation assumes fairly basic properties (such as spherical symmetry and a simple NFW model for the dark matter) that may not hold in the case of mergers. Furthermore, compared to the $\beta$-model, we make substantial improvements to fitting the central regions of the relaxed clusters, primarily due to one parameter, $\Gamma$.

\subsection{Extensions of the Model}\label{s:modeltests}
The results of the preceding sections have indicated that our model is in fact most strongly dependent on the parameter $\Gamma$, which controls the shape of the profile. Accordingly, it is possible to use this model as a one-parameter family of profiles by fixing $s$ to some reasonable value (for instance, for these low-redshift clusters, $s\approx 2$-$3\;\mathrm{Mpc}$ is acceptable since $s\sim{r_{vir}}\sim2r_{500}$) and also fixing $f_g=\Omega_{m}/\Omega_{DM}(\Omega_b/\Omega_m-f_{*})$, where $f_{*}$ is the fraction of baryonic matter in stars and for which values can be estimated independently based on optical/infrared data.

Our model, having been derived from the NFW profile, is of course dependent upon the NFW parameters $r_s$ and $\delta_c$, for which $r_{\Delta}$ and corresponding $c_{\Delta}$ (where $\Delta$ indicates the overdensity with respect to the critical density) should be determined. These can be obtained through other means, such as weak lensing. However, one can also leave these parameters free in the fit and obtain a model with the three free parameters ($\Gamma$, $r_s$, $\delta_c$) instead of ($\Gamma$, $f_g$, $s$). In any case, the model has significant freedom and yet enough simplicity to provide an attractive alternative to the $\beta$-model.

\section{Discussion}
By imposing only three simple cosmological and physical conditions that a gas density profile must satisfy, we have derived a family of profiles, characterized primarily by a single parameter $\Gamma$, that provide a good description of observational data over a wide range of radii. Our model can be used with 1-3 free parameters, each of which possesses a physical interpretation -- $\Gamma$ and $s$ come from the condition that there exist a virial shock on the outskirts of the cluster and represent its strength (relative to the dark matter jump) and location, respectively, while $f_g$ is a gas fraction parameter that fixes how much of the baryonic matter of the cluster is in the gas phase within $s$. However, upon comparison with X-ray data, we find that observations favor low values of $\Gamma$, the main parameter controlling the shape of the profile, which in turn leads to the $s$ parameter being poorly constrained. Additionally, average values of $f_g$ can be estimated from the literature, and fixing both it and $s$ to reasonable values leaves us with a one-parameter model. 

The derivation of our model assumes that the dark matter distribution internal to $s$ follows an NFW profile; accordingly, we have additional NFW parameters in our model. We do not necessarily count these as free parameters, since for many clusters they can be obtained from other data sets. However, in lieu of additional information, one can also consider an alternative model with the three parameters ($\Gamma$, $r_s$, $\delta_c$) and $s$ and $f_g$ fixed, as discussed above. 

Upon comparison with data, we find that our model provides enhanced flexibility over the $\beta$-model in the central regions of clusters, which can feature either prominent cores or cusps, while not introducing more parameters. Our model thus does not require an arbitrary radius cutoff, which is often necessary for modeling with the $\beta$ profile. We find satisfactory agreement between our model and the data for both relaxed clusters and mergers, although as a whole our model is a better description for relaxed systems. This is understandable, since mergers can violate some of the assumptions that we make. 

Overall, our family of functions provides a simple but accurate model of the galaxy cluster gas density distribution. We anticipate that it will have applications in both observational and theoretical work. From observations, we await confirmation of the virial shock, whose strength and location will provide an evaluation of our model parameters. On the theoretical side, our model can be used as a simple description of a wide range of cluster gas morphologies in numerical simulations. Furthermore, the prescription we present herein is not limited to the model that we present; in addition to the other families of profiles that we derived in $\S$\ref{s:method}, there are numerous extensions and modifications of our conditions that can be imposed to generate yet other models. Future efforts may build upon this framework to introduce additional physics of galaxy clusters into the modeling of their gas distributions.

\acknowledgments
We would like to thank Alexey Vikhlinin for very helpful discussions and providing the data used in $\S$\ref{s:obscomp}. We also greatly appreciate Bill Forman's thoughtful comments on a draft of this paper, and thank Andrey Kravtsov for bringing to our attention the recent papers on dark matter jumps. A.P. is supported by the National Science Foundation Graduate Research Fellowship under Grant No. DGE-1144152. This work was also supported by NSF Grant No. AST-1312034.

\end{document}